\documentclass[prl,aps,twocolumn,showpacs,groupauthors]{revtex4}

\usepackage{graphicx}
\usepackage{amsmath,amsfonts,amsthm}
\usepackage{bbm,mathrsfs}

\begin{document}

\title{A Deterministic and Nondestructively-Verifiable Photon Number Source}
\author{JM Geremia}
\email{jgeremia@unm.edu}
\affiliation{
\href{http://qmc.phys.unm.edu/}{Quantum Measurement \& Control}, 
\href{http://www.unm.edu}{The University of New Mexico}, Albuquerque, New Mexico 87131 USA}
\date{\today}
\begin{abstract}
We present a deterministic approach based on continuous measurement and real-time quantum feedback control to prepare arbitrary photon number states of a cavity mode.  The procedure passively monitors the number state actually achieved in each feedback stabilized measurement trajectory, thus providing a nondestructively verifiable photon source.   The feasibility of a possible cavity QED implementation in the many-atom good-cavity coupling regime is analyzed. 
\end{abstract}
\maketitle

Practical laboratory schemes for generating photon number states are highly sought-after for the enabling role these states play in quantum information science.  For example, single-photons  provide a primitive building block for quantum cryptographic systems, communication networks, and linear optical quantum computation \cite{Briegel2000,Knill2001}.  Multi-photon number states are desired for the future role they will play in sub-shotnoise quantum metrology and as a reagent for synthesizing complex nonclassical fields such as optical Sch\"{o}dinger cat states \cite{Brown2003}.

With this gathering of information-theoretic applications in mind, one can compile a list of those features that would be desired in an ideal photon source: (1) \textit{determinism}, meaning that the target photon number is produced with high probability in every state-preparation shot, (2) \textit{verifiability}, meaning that the number of photons actually generated can be diagnosed nondestructively in every shot, and (3) \textit{extendability}, meaning that a single device can vary the target photon number from shot to shot.  Different applications benefit from these features in different ways.   For instance, real-time verifiability (a true heralded single photon) is likely to be the highest priority for cryptographic security.  Meanwhile, determinism offers a photon source that is efficient for quantum communication and computation.  Ideally, a single physical device could meet all three objectives simultaneously.

To date, quality single photons have been produced from trapped-atom cavity QED experiments \cite{McKeever2004}, quantum dots \cite{Michler2000}, ballistic-atom cavity QED \cite{Kuhn2002}, and collective excitations of an atomic ensemble \cite{Chou2004}.  A theoretical approach for extending cavity QED systems to higher photon numbers by generalizing atomic dark-states to the symmetric group of $N$ atoms has been proposed \cite{Brown2003}.  However, trapped-atom cavity QED sources, while highly deterministic, require the experimentally-demanding strong-coupling regime for \textit{exactly} $N$ intra-cavity atoms to achieve an $N$-photon number state \cite{Brown2003}.  Varying the photon (atom) number in back-to-back shots in this regime is likely to be difficult.  It is not yet understood whether quantum dot, ballistic atom cavity QED, or atomic ensemble schemes can be reasonably extended to higher photon numbers.  None of these procedures are inherently single-shot verifiable. 

\begin{figure}[b]
\begin{center}
\includegraphics{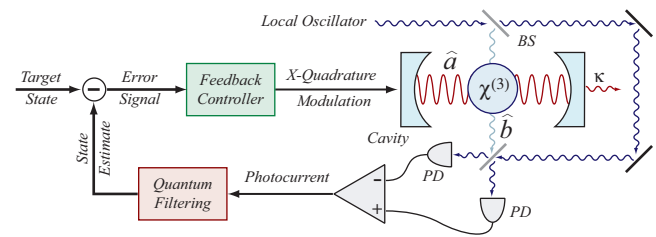}\end{center}
\vspace{-5mm}
\caption{(color online) Schematic of a feedback stabilized continuous measurement of cavity photon number for the purpose of preparing deterministic cavity photon number states. \label{Figure::Schematic}} 
\vspace{-1mm}
\end{figure}

In this paper, we present a photon number source that is simultaneously deterministic, intrinsically verifiable, and naturally capable of producing arbitrary number states.  Our procedure is based on a continuous indirect measurement \cite{Braginski1992} of photon number  \cite{Haroche1999} embedded within a real-time quantum feedback control loop \cite{Wiseman1994,Geremia2004a}.   Basic quantum mechanics specifies that conditioning on the outcome of a photon number measurement reduces the state of the field to a measurement eigenstate (or at least an approximate eigenstate in practice) \cite{Milburn1984}.   Quantum feedback renders this state reduction process deterministic by actively stabilizing the measurement outcome to the target eigenstate with arbitrarily high probability \cite{vanHandel2004}.   With a properly designed feedback policy, any possible measurement outcome (number eigenstate) is a viable candidate for stabilization in every distinct measurement trajectory.  And, since the state preparation is performed by a nondestructive quantum measurement, passive single-shot verifiability is an unavoidable fringe benefit of the photon source's own internal anatomy.

\begin{figure*}
\begin{center}\includegraphics{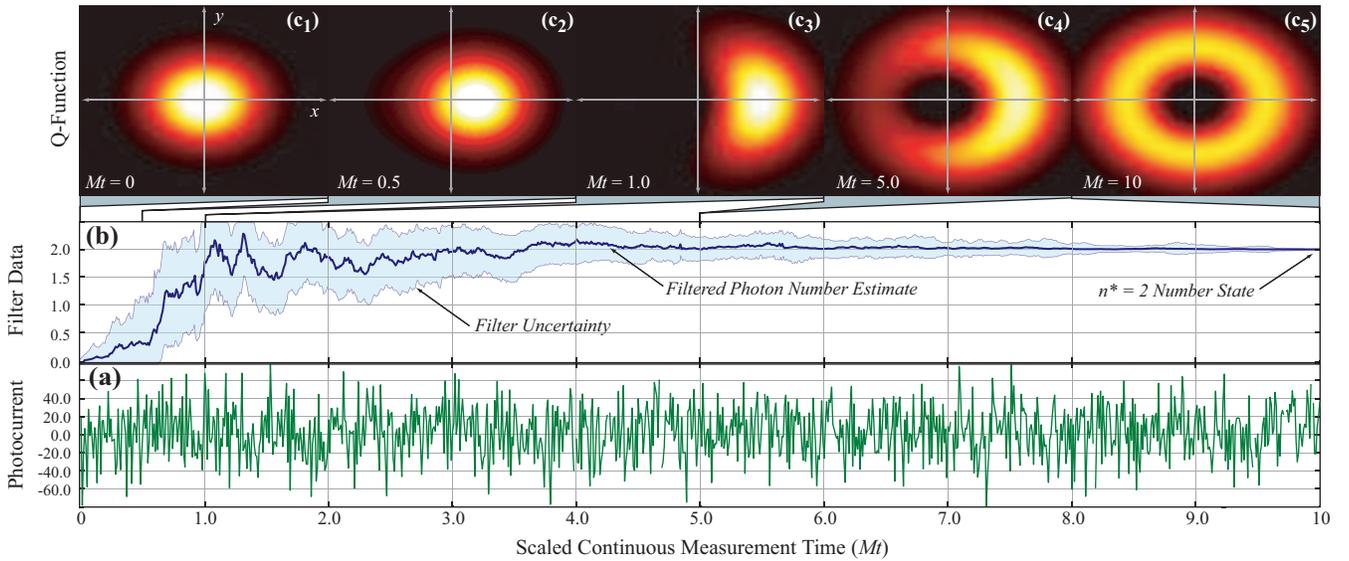}\end{center}
\vspace{-5mm}
\caption{(color online) A simulated photon number measurement trajectory ($M=10$, $G=20$, $\kappa=0$, $\eta=1$) where filtering the continuous photocurrent (a) provides an optimal real-time estimate (b) of the cavity photon number used to drive the measurement to a deterministic outcome via feedback. The evolving cavity mode $Q$-Function (c) illustrates the process.  \label{Figure::FeedbackTrajectory}} 
\end{figure*}

Figure \ref{Figure::Schematic} depicts a schematic of the feedback-stabilized source whose stability and experimental viability we analyze here.   A continuous measurement of photon number, $n$, in a single cavity mode, $\hat{a}$, is implemented by coupling that cavity mode to an auxiliary probe field, $\hat{b}$, via the cross-Kerr nonlinear scattering Hamiltonian, $\hat{H}_\mathrm{int} = \hbar \chi \hat{a}^\dagger \hat{a} \hat{b}^\dagger \hat{b}$ ($\chi$ is the strength of the Kerr nonlinearity).  For a coherent state probe, the mode-coupling interaction induces a phase-shift that is proportional to the cavity photon number, $n$ \cite{Imoto1985}.  Thus, observing the probe's phase provides an indirect measurement of $n$.  

Such a measurement can be implemented, as in Fig.\ \ref{Figure::Schematic}, by placing the Kerr nonlinearity in one arm of a Mach-Zender interferometer and then performing balanced homodyne detection on the forward-scattered probe field.  The output of the homodyne detectors in this configuration is given by the continuous photocurrent \cite{Wiseman1994,Doherty1999},
\begin{equation} \label{Equation::Photocurrent}
	d y_t  = 2 \eta \sqrt{M}\, n\, dt + \sqrt{\eta} \, dW_t ,
\end{equation}
where $\eta$ is the quantum efficiency of the photodetectors, $M$ is a rate referred to as the \textit{measurement strength} (described in detail below), and the $dW_t$ are Gaussian stochastic increments  that reflect quantum noise in the continuous measurement.   

Prior to conducting any feedback, these quantum fluctuations must be filtered from the photocurrent to obtain the continuous measurement outcome.  This means that an optimal time-dependent estimate of the cavity photon number must be derived from the photocurrent, $d y_t$ \cite{Doherty1999,Geremia2004a}.  The problem of extracting such an estimate is the subject of \textit{quantum filtering theory} \cite{Belavkin1999,vanHandel2004}, a field that combines elements from classical signal processing and stochastic analysis with the theory of continuously observed open quantum systems.  

Quantum filtering is conducted by propagating the cavity state, $\hat\rho$, according to a master equation,
\begin{eqnarray} \label{Equation::SME} 
	d \hat\rho_t & = & -i [ \hat{H}_t^\mathrm{(fb)}, \hat\rho_t ] \, dt + M \mathcal{D}[\hat{n}]\hat\rho_t \, dt
	 + \kappa \mathcal{D}[ \hat{a} ] \hat\rho_t \, dt \nonumber \\
	 & & + \sqrt{M} \mathcal{H} [ \hat{n} ] \hat\rho_t \left( dy_t  - 2 \eta\sqrt{M} \mathrm{tr}[ \hat{n}
	 	\hat\rho_t  ] \, dt \right) ,  
\end{eqnarray} 
that conditions it on the information provided by the accumulating measurement data, $d y_t$.  Here, $\mathcal{D}[\hat{r}]\hat\rho_t \equiv \hat{r}\hat\rho_t\hat{r}^\dagger - \frac{1}{2}( \hat{r}^\dagger\hat{r}\hat\rho_t + \hat\rho_t \hat{r}^\dagger\hat{r} )$ and $\mathcal{H}[\hat{r}]\hat\rho_t \equiv \hat{r} \hat\rho_t + \hat\rho_t \hat{r}^\dagger - \mathrm{tr}[ (\hat{r} + \hat{r}^\dagger) \hat\rho_t ] \hat\rho_t$.   The first term in the master equation describes any Hamiltonian driving performed on the system, the second term describes decoherence caused by coupling the cavity mode to the probe, the third term reflects cavity decay through the mirrors (with decay rate $\kappa$) and the final term conditions the state on the measurement via the innovation process, $dy_t - 2 \eta \sqrt{M} \langle \hat{n}\rangle_t dt$.

The optimal photon number estimate at time $t$ is obtained from the continuously conditioned cavity state as $\langle \hat{n} \rangle_t \equiv \mathrm{tr}[ \hat{n} \hat\rho_t]$.  Eq.\ (\ref{Equation::SME}) thus implicitly provides the crucial feedback ingredient known as the \textit{error signal},
\begin{equation}
	e_t = n^\star  - \langle \hat{n} \rangle_t,
\end{equation}
computed as the deviation of the estimated photon number from the target, $n^\star$.   

Feedback can then be performed by driving the cavity in response to the continuous error signal.   Here, we will consider a feedback control policy,
\begin{equation} \label{Equation::FeedbackLaw}
	\hat{H}_t^\mathrm{(fb)} = G \,e_t \, \hat{X},   
\end{equation}
that drives the cavity amplitude quadrature, $\hat{X} \equiv \frac{1}{2} ( \hat{a} + \hat{a}^\dagger)$, in proportion to the error signal with DC loop-gain, $G$.   This feedback policy has the satisfying intuitive interpretation that the controller will increase the amplitude of the intra-cavity field when its current estimate of the cavity photon number, $\langle \hat{n} \rangle_t$, is below that of the target, $n^\star$, and \textit{vice versa}.  

The simulated quantum feedback trajectory in Fig.\ \ref{Figure::FeedbackTrajectory} illustrates the salient features of our photon source for the idealized situation in which $\eta=1$ and $\kappa=0$ (more realistic simulations are described below).  The measurement begins with the cavity mode in the vacuum state with the objective to prepare an $n^\star=2$ number state.  The homodyne photocurrent, $y_t$ in Fig.\ \ref{Figure::FeedbackTrajectory}(a), is clearly swamped with quantum noise.  However, by propagating the quantum filtering equation subject to $y_t$ as more data becomes available, the controller extracts its optimal real-time estimate of the cavity photon number [Fig.\ \ref{Figure::FeedbackTrajectory}(b)].  Uncertainty in the estimate [shaded region in Fig.\ \ref{Figure::FeedbackTrajectory}(b)] is gradually reduced by the measurement.  

\begin{figure}[b]
\vspace{-2mm}
\begin{center}
\includegraphics{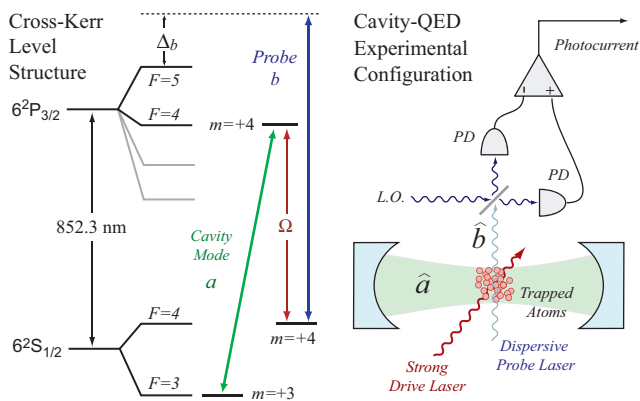}
\end{center}
\vspace{-6mm}
\caption{Cavity QED implementation of the continuous photon number measurement. \label{Figure::LevelStructure}}
\end{figure}

Figures \ref{Figure::FeedbackTrajectory}(c$_1$-c$_5$) highlight the feedback-mediated progression to the target eigenstate by depicting the cavity $Q$-Function as a function of time ($Q_t( \alpha ) = \frac{1}{\pi} \langle \alpha | \hat\rho_t | \alpha \rangle$ is a quasiprobability distribution parameterized by the coherent state amplitude, $\alpha = x + i y \in \mathbbm{C}$).  Beginning from the vacuum $Q$-Function, $Q_0( \alpha ) = \frac{1}{\pi} \exp\left(- \frac{1}{4}|\alpha|^2\right)$ in Fig.\ \ref{Figure::FeedbackTrajectory}(c$_1$), the quantum filter rapidly finds that the cavity photon number is small relative to $n^*=2$.  Feedback consequently displaces the cavity mode toward a coherent state with $\langle \hat{n}\rangle \sim 2$ by driving its amplitude quadrature, seen in Fig.\ \ref{Figure::FeedbackTrajectory}(c$_2$) at time $Mt\sim 0.1$.  As the cavity state is further conditioned on the measurement data, the photon number uncertainty, $\langle \Delta \hat{n} \rangle$, begins to decrease at the expense of phase uncertainty [Fig.\ \ref{Figure::FeedbackTrajectory}(c$_3$)], eventually producing a heavily number-squeezed state [Fig.\ \ref{Figure::FeedbackTrajectory}(c$_4$) at $Mt \sim 5$].  The target $n^*=2$ eigenstate--- with its tell-tale phase-delocalized $Q$-Function--- is ultimately achieved [Fig.\ \ref{Figure::FeedbackTrajectory}(c$_5$) at $Mt=10$]. 

In addition to this feedback policy, we also require the means to implement the cross-Kerr Hamiltonian underlying the continuous measurement \cite{Imoto1985}.  A suitable such interaction is provided by a cavity-enhanced analog of the atomic dark-state level structure originally proposed for generating giant Kerr nonlinearities in free space \cite{Schmidt1996}.   We consider a sample of $N$ intracavity atoms with the hyperfine level structure depicted in Fig.\ \ref{Figure::LevelStructure}.  In this scheme, two non-radiative hyperfine stretched states are coherently coupled via a two-photon transition that involves both the cavity mode as well as a strong drive laser. 

A photocurrent of the form in Eq.\ (\ref{Equation::Photocurrent}) is obtained by dispersively coupling the probe field to the atomic hyperfine structure.  In the limit where atomic motion can be neglected, as would be the case for trapped intra-cavity atoms, we find the measurement strength, $M$, to be
\begin{equation} \label{Equation::QEDM}
	M = \frac{P_b}{\hbar \omega_b}  \left[ \frac{3 N \Gamma \lambda_b^2}
	{4 \pi^2 r^2 \Delta_b} \left( \frac{g_0^2}{g_0^2+\Omega^2} \right) \right]^2,
\end{equation}
where $P_b$ is the optical power of the dispersive probe, $\omega_b$ (or $\lambda_b$) is the probe frequency (or wavelength), $\Delta_b$ is the probe detuning, $\Gamma$ is the atomic spontaneous emission rate, $r$ is the radius of the atomic sample, $g_0$ is the single photon cavity coupling rate, and $\Omega=\Gamma \sqrt{I/2 I_\mathrm{sat}}$ is the Rabi frequency associated with the drive laser.

\begin{table}[t]
\vspace{-3mm}
\caption{Simulation parameters used to analyze experimental feasibility of a cavity QED feedback implementation. \label{Table::Parameters}}
\begin{tabular*}{3.4in}{c|@{\extracolsep{\fill}}ccc@{\hspace{4mm}}} \hline\hline
	Parameter			& Symbol 			&	Value	& Units \\ \hline
	Probe Power			& $P_b$ 			&	1		& $\mu$W \\
	Probe Wavelength		& $\lambda_b$		&	852.35      & nm \\
	Probe Detuning		& $\Delta_b$		& 	2	         & GHz \\
	Cavity Decay Rate		& $\kappa$		&	12		& kHz \\
	Cavity Coupling Rate 	& $g_0$			&	200	         & kHz \\
	Atom Sample Radius	& $r$			&	110	         & $\mu$m \\
	Drive Laser Intensity		& $I$				&	1/4		& $I_\mathrm{sat}$ \\
	Feedback DC Gain		& $G$			&	20	         & dB \\
	Detector Efficiency		& $\eta$			& 	80		& \%  \\ \hline\hline
\end{tabular*}
\end{table}

\begin{figure*}
\begin{center}\includegraphics{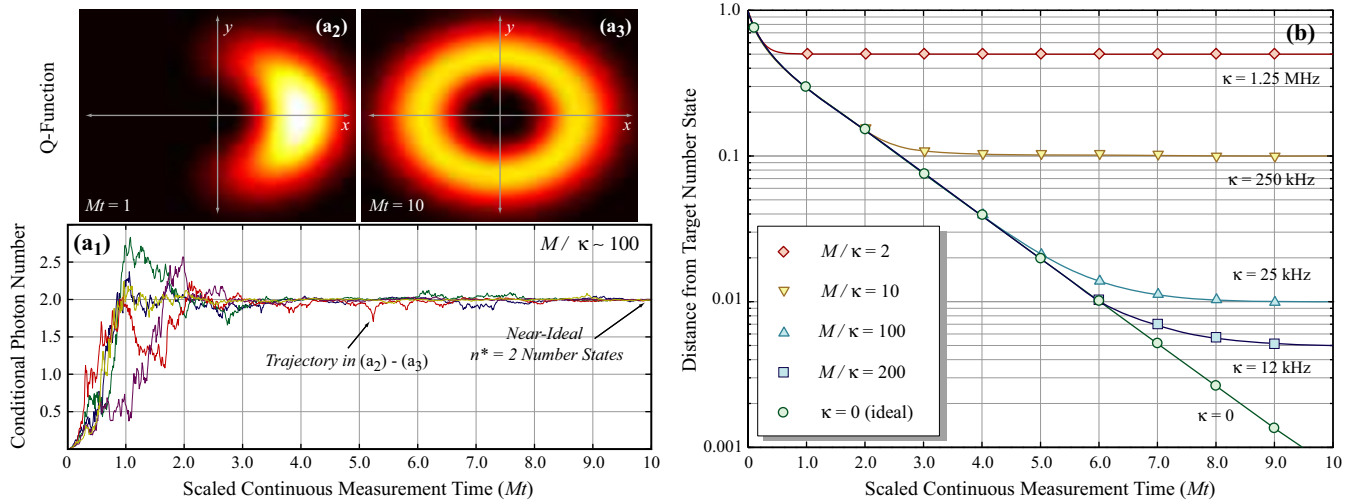}\end{center}
\vspace{-3mm}
\caption{(color online) Simulated feedback-stabilized photon number measurement trajectories (a) generated using the parameters listed in Table \ref{Table::Parameters} and feedback stability (b) for different cavity decay rates (solid lines depict theoretical results). \label{Figure::DecayTrajectory}} 
\vspace{-2mm}
\end{figure*}

In order to analyze the performance of this measurement scheme under realistic conditions, we envisage trapping $N \sim 1\times 10^6$ Cs atoms within the mode volume of a $L \sim 4$ cm Fabry-Perot cavity with mirror finesse, $\mathcal{F} \sim 3 \times 10^5$.  Measurement parameters and experimental values derived from these cavity properties are listed in Table \ref{Table::Parameters}.  Using Eq.\ (\ref{Equation::QEDM}) we find the measurement strength to be $M \sim 2.5$ MHz compared to a cavity decay rate of $\kappa \sim 12$ kHz.   These parameters correspond to the many-atom cavity QED strong-coupling regime, $\sqrt{N} g_0^2 / \Gamma \kappa \gg 1$.  Given that the total atomic scattering rate for these parameters is $\gamma_\mathrm{s} \sim 2.5$ kHz,  spontaneous emission due to the probe can be reasonably neglected.

Figure \ref{Figure::DecayTrajectory} demonstrates the level of performance that we expect from the feedback-stabilized measurement given the parameters in Table \ref{Table::Parameters}.   Five typical simulated measurement trajectories are shown in  Fig.\ \ref{Figure::DecayTrajectory}(a$_1$).  Comparing the $Q$-functions in Figs.\ \ref{Figure::DecayTrajectory}(a$_2$ -- a$_3$) with those in Fig.\ \ref{Figure::FeedbackTrajectory} suggests that there is little qualitative difference in this measurement relative to the ideal $\kappa=0$ case.   In general, we expect that $M \gg \kappa$ will be needed such that the measurement can reduce the cavity mode to a good approximate number state prior to appreciable decay.

A quantitative analysis of the feedback stability was conducted by computing the distance of the cavity state from the target eigenstate as a function of time.  To do so, we employed the following distance metric,
\begin{equation} \label{Equation::VFunctional}
	\mathscr{D}[Q_t] = 1 - \frac{1}{\pi 4^{(n^\star+1)} n^\star!} \int_{\mathbbm{C}}
		|\alpha|^{2n^\star} e^{-\frac{1}{4} |\alpha|^2} Q_t d \alpha .
\end{equation}
Note that $\mathscr{D}$ is a scalar functional from the space of $Q$-functions into the real numbers between zero and one, $\mathscr{D}[\cdot] : L_2( \mathbbm{C} ) \rightarrow \mathbbm{R}_{[0,1]}$.  Moreover, $\mathscr{D}$ assumes the value $\mathscr{D}[Q_t]=0$ only when the cavity is in the target number state.   It assumes its maximum value, $\mathscr{D}[Q_t]=1$, when the cavity is in any other number state and smoothly interpolates otherwise.  We find that the distance between the cavity state and the target eigenstate, as measured by $\mathscr{D}$, strictly decreases in expectation with time, $d \mathbbm{E}[ \mathscr{D}[Q_t]] \le 0$.  Figure \ref{Figure::DecayTrajectory}(b) illustrates the time evolution of the distance measure averaged over $10^5$ trajectories for different cavity decay rates.  The experimental parameters in Table \ref{Table::Parameters} correspond to $M/\kappa \sim 200$.  As expected, the quality of the state preparation is degraded for high cavity decay.   While an analytic stability proof does not appear to exist for $\kappa > 0$, this numerical evidence for feedback stability when $M\gg \kappa$ is substantial.

Given our findings, it appears that quantum feedback stabilization of a continuous cavity photon number does indeed provide a deterministic photon source that is real-time verifiable and capable of producing arbitrary number states.   Our feasibility assessment for a cavity QED implementation of this scheme using currently available laboratory technology is favorable.   Additionally, we anticipate that our procedure will be robust to reasonable uncertainty in the intra-cavity atom number as each individual atom is only weakly coupled to the cavity.  Our procedure therefore avoids hidden deterministic prerequisites, such as having to trap exactly $N^*$ atoms to produce the number state with $n^*=N^*$ photons \cite{McKeever2004,Brown2003}.  

In light of these results, we anticipate that real-time feedback-stabilized quantum state reduction will play an increasingly powerful role in meeting the state preparation objectives demanded by precision quantum information theoretic applications.  The author would like to thank Ivan Deutsch and Ramon van Handel for comments.  Additional information can be found at our group website, \url{http://qmc.phys.unm.edu}.


\vspace{-2mm}
\begin{thebibliography}{18}
\expandafter\ifx\csname natexlab\endcsname\relax\def\natexlab#1{#1}\fi
\expandafter\ifx\csname bibnamefont\endcsname\relax
  \def\bibnamefont#1{#1}\fi
\expandafter\ifx\csname bibfnamefont\endcsname\relax
  \def\bibfnamefont#1{#1}\fi
\expandafter\ifx\csname citenamefont\endcsname\relax
  \def\citenamefont#1{#1}\fi
\expandafter\ifx\csname url\endcsname\relax
  \def\url#1{\texttt{#1}}\fi
\expandafter\ifx\csname urlprefix\endcsname\relax\def\urlprefix{URL }\fi
\providecommand{\bibinfo}[2]{#2}
\providecommand{\eprint}[2][]{\url{#2}}

\bibitem[{\citenamefont{Briegel et~al.}(2000)\citenamefont{Briegel, van Enk,
  Cirac, and Zoller}}]{Briegel2000}
\bibinfo{author}{\bibfnamefont{H.-J.} \bibnamefont{Briegel}},
  \bibinfo{author}{\bibfnamefont{et~al.}},  in
  \emph{\bibinfo{booktitle}{The Physics of Quantum Information}}, edited by
  \bibinfo{editor}{\bibfnamefont{D.}~\bibnamefont{Bouwmeester}},
  \bibinfo{editor}{\bibfnamefont{A.}~\bibnamefont{Ekert}}, \bibnamefont{and}
  \bibinfo{editor}{\bibfnamefont{A.}~\bibnamefont{Zeilinger}}
  (\bibinfo{publisher}{Springer}, \bibinfo{address}{Berlin},
  \bibinfo{year}{2000}), p. \bibinfo{pages}{192}.

\bibitem[{\citenamefont{Knill et~al.}(2001)\citenamefont{Knill, Laflamme, and
  Milburn}}]{Knill2001}
\bibinfo{author}{\bibfnamefont{E.}~\bibnamefont{Knill}},
  \bibinfo{author}{\bibfnamefont{R.}~\bibnamefont{Laflamme}}, \bibnamefont{and}
  \bibinfo{author}{\bibfnamefont{G.}~\bibnamefont{Milburn}},
  \bibinfo{journal}{Nature} \textbf{\bibinfo{volume}{409}}, \bibinfo{pages}{46}
  (\bibinfo{year}{2001}).

\bibitem[{\citenamefont{Brown et~al.}(2003)\citenamefont{Brown, Dani,
  Stamper-Kurn, and Whaley}}]{Brown2003}
\bibinfo{author}{\bibfnamefont{K.}~\bibnamefont{Brown}},
  \bibinfo{author}{\bibfnamefont{K.}~\bibnamefont{Dani}},
  \bibinfo{author}{\bibfnamefont{D.}~\bibnamefont{Stamper-Kurn}},
  \bibnamefont{and} \bibinfo{author}{\bibfnamefont{K.}~\bibnamefont{Whaley}},
  \bibinfo{journal}{Phys. Rev. A} \textbf{\bibinfo{volume}{67}},
  \bibinfo{pages}{043818} (\bibinfo{year}{2003}).

\bibitem[{\citenamefont{McKeever et~al.}(2004)\citenamefont{McKeever, Boca,
  Boozer, Miller, Buck, Kuzmich, and Kimble}}]{McKeever2004}
\bibinfo{author}{\bibfnamefont{J.}~\bibnamefont{McKeever}},
  \bibinfo{author}{et~al.},
  \bibinfo{journal}{Science} \textbf{\bibinfo{volume}{303}},
  \bibinfo{pages}{1992} (\bibinfo{year}{2004}).

\bibitem[{\citenamefont{Michler et~al.}(2000)\citenamefont{Michler, Kiraz,
  Cecher, Schoenfeld, Petroff, Zhang, Hu, and Imamoglu}}]{Michler2000}
\bibinfo{author}{\bibfnamefont{P.}~\bibnamefont{Michler}},
  \bibinfo{author}{et~al.},
  \bibinfo{journal}{Science} \textbf{\bibinfo{volume}{290}},
  \bibinfo{pages}{2282} (\bibinfo{year}{2000});
\bibinfo{author}{\bibfnamefont{M.}~\bibnamefont{Pelton}},
  \bibinfo{author}{et~al.},
  \bibinfo{journal}{Phys. Rev. Lett.} \textbf{\bibinfo{volume}{89}},
  \bibinfo{pages}{233602} (\bibinfo{year}{2002}).

\bibitem[{\citenamefont{Kuhn et~al.}(2002)\citenamefont{Kuhn, Hennrich, and
  Rempe}}]{Kuhn2002}
\bibinfo{author}{\bibfnamefont{A.}~\bibnamefont{Kuhn}},
  \bibinfo{author}{\bibfnamefont{M.}~\bibnamefont{Hennrich}}, \bibnamefont{and}
  \bibinfo{author}{\bibfnamefont{G.}~\bibnamefont{Rempe}},
  \bibinfo{journal}{Phys. Rev. Lett.} \textbf{\bibinfo{volume}{89}},
  \bibinfo{pages}{067901} (\bibinfo{year}{2002}).

\bibitem[{\citenamefont{Chou et~al.}(2004)\citenamefont{Chou, Polyakov,
  Kuzmich, and Kimble}}]{Chou2004}
  \bibinfo{author}{\bibfnamefont{D.N.}~\bibnamefont{Matsukevich}},   \bibinfo{author}{et~al.},
  \bibinfo{journal}{Phys. Rev. Lett.} \textbf{\bibinfo{volume}{95}},
  \bibinfo{pages}{040405} (\bibinfo{year}{2005}); \bibinfo{author}{\bibfnamefont{C.}~\bibnamefont{Chou}},
  \bibinfo{author}{\bibfnamefont{S.}~\bibnamefont{Polyakov}},
  \bibinfo{author}{\bibfnamefont{A.}~\bibnamefont{Kuzmich}}, \bibnamefont{and}
  \bibinfo{author}{\bibfnamefont{H.}~\bibnamefont{Kimble}},
  \bibinfo{journal}{Phys. Rev. Lett.} \textbf{\bibinfo{volume}{92}},
  \bibinfo{pages}{213601} (\bibinfo{year}{2004}).

\bibitem[{\citenamefont{Braginski and Khalili}(1992)}]{Braginski1992}
\bibinfo{author}{\bibfnamefont{V.}~\bibnamefont{Braginski}} \bibnamefont{and}
  \bibinfo{author}{\bibfnamefont{F.}~\bibnamefont{Khalili}},
  \emph{\bibinfo{title}{Quantum Measurements}} (\bibinfo{publisher}{Cambridge
  University Press}, \bibinfo{year}{1992}).

\bibitem[{\citenamefont{Nogues et~al.}(1999)\citenamefont{Nogues,
  Rauschenbeutel, Osnaghi, Brune, Raimond, and Haroche}}]{Haroche1999}
\bibinfo{author}{\bibfnamefont{G.}~\bibnamefont{Nogues}},
  \bibinfo{author}{et~al.},
  \bibinfo{journal}{Nature} \textbf{\bibinfo{volume}{400}},
  \bibinfo{pages}{239} (\bibinfo{year}{1999}).

\bibitem[{\citenamefont{Wiseman and Milburn}(1994)}]{Wiseman1994}
\bibinfo{author}{\bibfnamefont{H.~M.} \bibnamefont{Wiseman}} \bibnamefont{and}
  \bibinfo{author}{\bibfnamefont{G.~J.} \bibnamefont{Milburn}},
  \bibinfo{journal}{Phys. Rev. A} \textbf{\bibinfo{volume}{49}},
  \bibinfo{pages}{4110} (\bibinfo{year}{1994}).

\bibitem[{\citenamefont{Geremia et~al.}(2004)\citenamefont{Geremia, Stockton,
  and Mabuchi}}]{Geremia2004a}
\bibinfo{author}{\bibfnamefont{J.}~\bibnamefont{Geremia}},
  \bibinfo{author}{\bibfnamefont{J.~K.} \bibnamefont{Stockton}},
  \bibnamefont{and} \bibinfo{author}{\bibfnamefont{H.}~\bibnamefont{Mabuchi}},
  \bibinfo{journal}{Science} \textbf{\bibinfo{volume}{304}},
  \bibinfo{pages}{270} (\bibinfo{year}{2004}).

\bibitem[{\citenamefont{Milburn and Walls}(1984)}]{Milburn1984}
\bibinfo{author}{\bibfnamefont{G.~J.} \bibnamefont{Milburn}} \bibnamefont{and}
  \bibinfo{author}{\bibfnamefont{D.}~\bibnamefont{Walls}},
  \bibinfo{journal}{Phys. Rev. A} \textbf{\bibinfo{volume}{30}},
  \bibinfo{pages}{56} (\bibinfo{year}{1984}).

\bibitem[{\citenamefont{van Handel et~al.}(2005)\citenamefont{van Handel,
  Stockton, and Mabuchi}}]{vanHandel2004}
\bibinfo{author}{\bibfnamefont{R.}~\bibnamefont{van Handel}},
  \bibinfo{author}{\bibfnamefont{J.~K.} \bibnamefont{Stockton}},
  \bibnamefont{and} \bibinfo{author}{\bibfnamefont{H.}~\bibnamefont{Mabuchi}},
  \bibinfo{journal}{IEEE Trans. Aut. Control} \textbf{\bibinfo{volume}{50}},
  \bibinfo{pages}{768} (\bibinfo{year}{2005}).

\bibitem[{\citenamefont{Imoto et~al.}(1985)\citenamefont{Imoto, haus, and
  Yamamoto}}]{Imoto1985}
\bibinfo{author}{\bibfnamefont{N.}~\bibnamefont{Imoto}},
  \bibinfo{author}{\bibfnamefont{H.}~\bibnamefont{haus}}, \bibnamefont{and}
  \bibinfo{author}{\bibfnamefont{Y.}~\bibnamefont{Yamamoto}},
  \bibinfo{journal}{Phys. Rev. A} \textbf{\bibinfo{volume}{32}},
  \bibinfo{pages}{2287} (\bibinfo{year}{1985}).

\bibitem[{\citenamefont{Doherty et~al.}(1999)\citenamefont{Doherty, Tan,
  Parkins, and Walls}}]{Doherty1999}
\bibinfo{author}{\bibfnamefont{A.~C.} \bibnamefont{Doherty}},
  \bibinfo{author}{\bibfnamefont{S.~M.} \bibnamefont{Tan}},
  \bibinfo{author}{\bibfnamefont{A.~S.} \bibnamefont{Parkins}},
  \bibnamefont{and} \bibinfo{author}{\bibfnamefont{D.~F.} \bibnamefont{Walls}},
  \bibinfo{journal}{Phys. Rev. A} \textbf{\bibinfo{volume}{60}},
  \bibinfo{pages}{2380} (\bibinfo{year}{1999}).

\bibitem[{\citenamefont{Belavkin}(1999)}]{Belavkin1999}
\bibinfo{author}{\bibfnamefont{V.}~\bibnamefont{Belavkin}},
  \bibinfo{journal}{Rep. on Math. Phys.} \textbf{\bibinfo{volume}{43}},
  \bibinfo{pages}{405} (\bibinfo{year}{1999}).

\bibitem[{\citenamefont{Schmidt and Imamoglu}(1996)}]{Schmidt1996}
\bibinfo{author}{\bibfnamefont{H.}~\bibnamefont{Schmidt}} \bibnamefont{and}
  \bibinfo{author}{\bibfnamefont{A.}~\bibnamefont{Imamoglu}},
  \bibinfo{journal}{Optics Letters} \textbf{\bibinfo{volume}{21}},
  \bibinfo{pages}{1936} (\bibinfo{year}{1996}).

\end{thebibliography}

\end{document}